\documentclass{ptapap}
\usepackage{amsmath}
\usepackage{amssymb}
\usepackage{color}

\usepackage{soul}

\author{Mohammad-Hassan Naddaf}[CFT, CAMK]
\author{Bo\.zena Czerny}[CFT]
\affil[CFT]{Center for Theoretical Physics,
  Lotnikow 32/46, 02--668 Warszawa, Poland}
\affil[CAMK]{Nicolaus Copernicus Astronomical Center,
  Bartycka 18, 00--716 Warszawa, Poland}
\title{Mass loss rate of accretion disk in FRADO}

\begin{document}

\maketitle

\begin{abstract}

We have developed the 2.5D version of the basic physically motivated 1D model of \citet{czerny2011}, i.e. Failed Radiatively Accelerated Dusty Outflow (FRADO) model. This model is based on the idea that radiation pressure acting on dust is responsible for the formation of the low ionized part of the Broad Line Region (BLR). Such radiation pressure is strong enough to form a fast outflow from the disk surface in the inner part of low ionized BLR. The outflow properties depend on the basic physical parameters, like black hole mass, Eddington ratio and gas metallicity. We here aim at estimating the disk mass loss rate due to this process, and comparing the results with outflows detected in Broad Absorption Line (BAL) quasars.

\end{abstract}

\section{Introduction}

There are many studies in hydro and non-hydro context reporting the presence of a fast funnel-shaped outflow of material from accretion disk triggered by the disk radiation pressure \citep[see e.g.][and the references therein]{proga2004, Risaliti2010, Nomura2020}; in line with the empirical picture predicted by \citet{elvis2000}. These models are based on the line-driving mechanism appropriate for the highly-ionized part of BLR located at a few hundred gravitational radii, $r_{\rm g}$. Interestingly, FRADO model also gives rise to formation of fast outflow from the disk, although it works at the basis of dust-driving mechanism and addresses the farther part of BLR located at around a few thousand $r_{\rm g}$ where the disk is cold enough for dust to develop. The basic 1D model \citep{czerny2011} could not catch the outflow feature, but the recently developed 2.5D enhanced version of model \citep{naddaf2021} predicted the development of such feature in the low-ionized part of BLR due to radiation pressure acting on dust. Notably, the predictions of the new version were recently confronted against the radius-luminosity relation \citep{naddaf2020} and the observed shape of emission lines \citep{naddafczerny2021}, hence making it an interesting model for further studies. As the outflows can generally cause the disk to lose mass which affect the accretion process, so it is worth to examine the strength of mass loss rate. Therefore, we intend in this work to estimate the mass loss rate from accretion disk in the low-ionized BLR based on 2.5D FRADO model. 

\section{Outflow stream}

In our first study of the dynamics of the low-ionized BLR in FRADO \citep{naddaf2021}, the black hole mass, and dust-to-gas ratio of BLR material was fixed at $10^8 M_{\odot}$, and $0.005$, respectively; in which the dust-to-gas mass ratio is equivalent to a medium with solar metallicity. The funnel-shaped stream feature was seen there only for the case of high accretion rate, i.e. one Eddington. The stream was relatively narrow, with the width of \emph{escaping zone} (the radial region of the disk from which the launched clouds escape) equal to $\Delta R=51~r_{\rm g}$, starting at the distance of $R_{\rm in}^{\rm s} = 5650~r_{\rm g}$ from the central black hole. Later, with an enhancement of the dust-to-gas ratio to a more typical 5 times solar value, we observed formation of the same feature even in sources with a lower accretion rate \citep{naddafczerny2021}. It was also seen that the width of \emph{escaping zone} increases with dust-to-gas ratio implying a larger mass loss rate from the underlying accretion disk. We here refer to our very recent simulations \citep[see][for more details]{naddafczerny2021} which cover a large grid of initial conditions i.e. black hole mass, accretion rate, and dust-to-gas ratio, based on which we examine the mass loss rate from the disk.

\begin{figure}
\includegraphics[width=\textwidth]{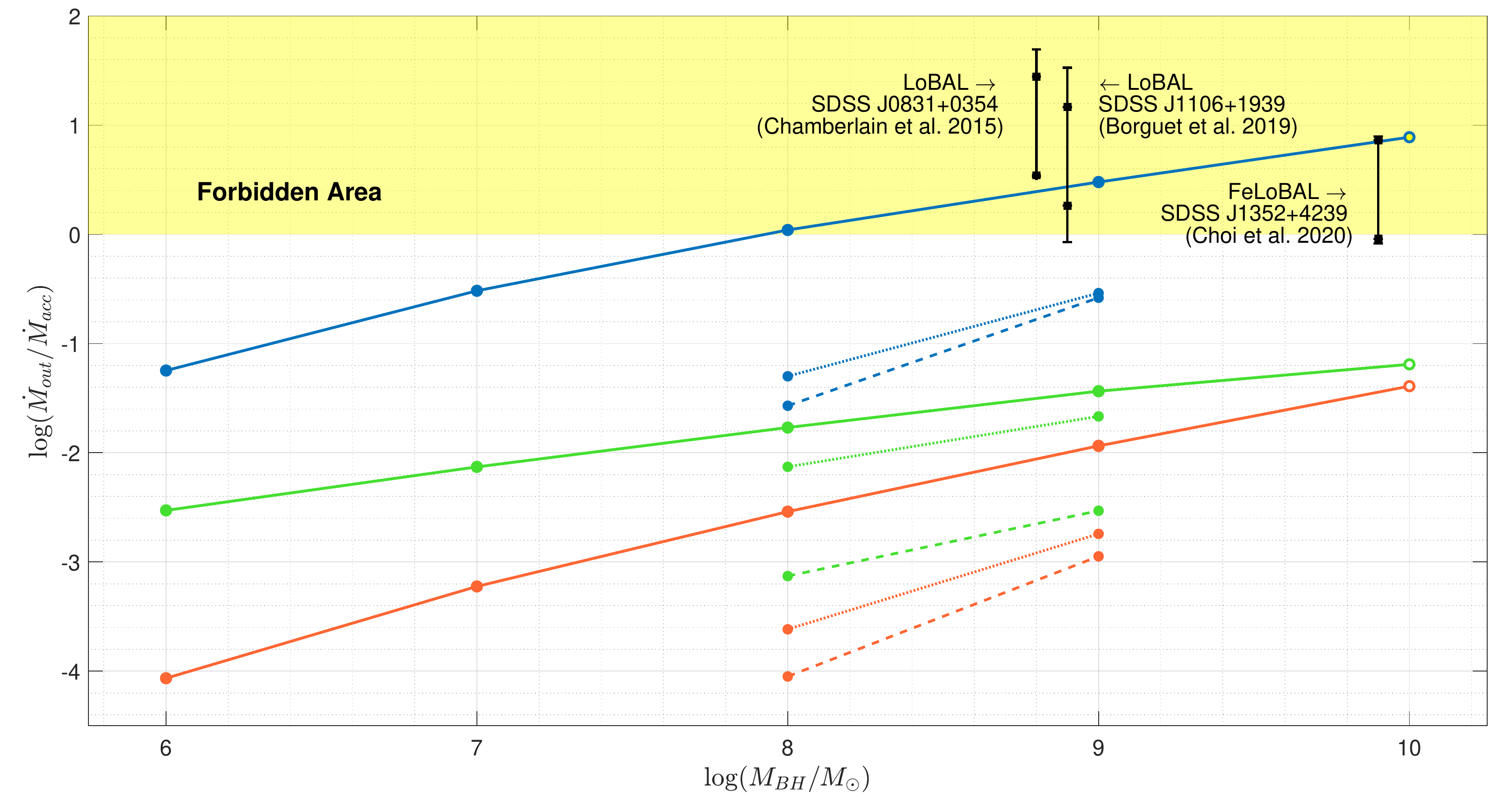}
\caption{Results obtained for $\dot M_{\rm out}^{\rm 1}$, $\dot M_{\rm out}^{\rm 2}$, and $\dot M_{\rm out}^{\rm 3}$ (divided by accretion rate) are color-coded with red, blue, and green, respectively. Solid, dotted, and dashed lines represent the [Eddington rate, metalicity] of [1, 5], [0.1, 5], and [1, 1], respectively. Hollow-circles are extrapolated values. Accretion efficiency is set to 0.1. Observational data for three BAL QSO near-Eddington sources are shown in black, with the upper/lower points corresponding to outflow solid angles of $0.08$/$0.01$, and with error bars included.}
\label{fig:massloss}
\end{figure}

In order to estimate the ejected mass from the disk within the \emph{escaping zone}, we follow three methods. The first two methods are adopted from \citet{czerny2017}. One based on applying the optically-thin approximation, with the momentum of radiation transferred to outflow, in stellar winds to accretion disks which yields
\begin{equation}
\dot M_{\rm out}^{\rm 1} =
\frac{3}{4} \dot M_{\rm acc} \sqrt{r_{\rm g}} ~\int^{R_{\rm out}^{\rm s}}_{R_{\rm in}^{\rm s}}
\left(\frac{1}{r}\right)^{\frac{3}{2}} ~ dr
\end{equation}
and the other one in which assuming multiple scatterings the whole energy is transferred to the outflow so
\begin{equation}
\dot M_{\rm out}^{\rm 2} = \frac{3}{2} \dot M_{\rm acc}~\int^{R_{\rm out}^{\rm s}}_{R_{\rm in}^{\rm s}}
\frac{1}{r} ~ dr
\end{equation}
and the third method as in \citet{naddaf2021} is based on assuming optically thin clouds at the time of launching that gives 
\begin{equation}
\dot M_{\rm out}^{\rm 3} =
\frac{\pi\ (R^{\rm s}_{\rm in}+R^{\rm s}_{\rm out}) ~ \Delta R\ m_{\rm p}}{t_{\rm exit} ~\sigma_{\rm Pl}}
\end{equation}
where $t_{\rm exit}$ is the time needed for the ejected material to leave the \emph{escaping zone}, $\sigma_{\rm Pl}$ is the Planck opacity at sublimation radius, $m_{\rm p}$ is the mass of proton, $R_{\rm in}^{\rm s}$/$R_{\rm out}^{\rm s}$ is the inner/outer radius of the \emph{escaping zone}, $\dot M_{\rm acc}$ is the disk accretion rate, and $\dot M_{\rm out}$ is the disk mass loss rate.

\section{Results and Conclusion}

The results are shown in figure \ref{fig:massloss}. As can be seen, the outflow rate increases with blackhole mass, accretion rate, and metallicity, with the role of metallicity stronger than accretion rate. Compared to the whole accretion rate for a given source, the values are small but not negligible, except for the optically-thick approximation (2nd method) comparable to the massive outflow seen in BAL quasars \citep{Borguet2013, chamberlain2015, choi2020}. But the material in the low ionized part of BLR is not fully but moderately thick, and also the disk-originated outflow rate must not exceed the the underlying disk accretion rate marked in yellow as forbidden area in the plot. Therefore, apart from a part of the observational data of BAL quasars located in the forbidden area, the rest also cannot be explained by the modeled stream of material, unless the dust content of BLR material is highly super solar around 10 times solar, or even higher. The other options, otherwise, that can override the forbidden area is that the huge amount of material outflowing in BAL quasars must be supplied from elsewhere rather than the disk itself; or the disk accretion rate at large radii, for example at the location of low ionized BLR, is indeed higher than the value at the innermost parts determining the bolometric luminosity. We will study this issue with further details in future.

\acknowledgements{The project was partially supported by NCN grant No. 2017/26/A/ST9/ 00756 (Maestro 9), and by MNiSW grant No. DIR/WK/2018/12.}

\bibliographystyle{ptapap}
\bibliography{naddaf}

\end{document}